# On some minimal characteristics in a model of a system of $N$ particles with interaction

Igor Pavlov

Abstract: For the well-known model of a system of $N$ particles with interaction (the $N$-body problem), we consider the spatial problem of finding the minimum of the function of the kinetic energy of a system on its phase space under conditions on its size and angular momentum. Based on the solution to this problem, we prove that the minimum possible kinetic energy of a system at the current value of its size can be achieved only on flat trajectories of the system. And under some natural additional conditions these trajectories are flat finite and periodic (elliptical) trajectories generated by flat central configurations. The solution to this problem also provides a simpler solution to a similar optimization dual problem of finding the minimum of the size of a system under conditions on its kinetic energy and angular momentum. This leads to a similar result that the minimum possible size of the system at the current value of its kinetic energy can also be achieved only on flat trajectories. And under some additional conditions these trajectories are also flat periodic elliptical trajectories generated by flat central configurations. Next, we consider the more complex spatial problem of finding local minima of the function of the kinetic energy of a system on its phase space at fixed values of the integrals of motion: angular momentum and total energy of a system. Based on the solution of this problem, we prove further that under some natural additional conditions the local minima of the system kinetic and potential energy functions at fixed values of these integrals of motion can be achieved only on flat periodic elliptical trajectories generated by some flat central configurations. And at the points of these local minima the minimum possible kinetic energy of the system at the current value of its size and the minimum possible size of the system at the current value of its kinetic energy are also achieved.

## 1. Introduction

Due to the complexity of this problem, in addition to the general problem, various special cases are also considered. An exact analytical solution was obtained only for the three-body problem in the form of well-known series of Sundman [1] (which, however, converge too slowly for practical use). In the general case, this problem has 10 known integrals ([1], [2], etc.), which is also not enough to solve when the number of particles in the system increases. Conditionally periodic solutions for the case when the masses of (N – 1) bodies are small compared to the mass of the central body were found in [3]. In [4], a spatial problem was considered in which $N_1$ particles with equal masses are located in a plane at the vertices of a regular $N_1$-gon and another $N_2$ particles with equal masses are located symmetrically on a line perpendicular to this plane. In [5], a spatial problem was studied in which N bodies of equal mass are initially located at the vertices of a polyhedron and another central body is located at the geometric center of the polyhedron. In [6], a situation is considered when N particles with equal masses are located at the vertices of a regular polyhedron and there is another particle with a low mass in the system. The spatial problem for the case of five bodies was also considered in [7].

A significant number of works are devoted to the problem of N bodies on a plane. In [8], symmetrical structures (central configurations) of the system are considered for the case of N = 4 particles. In [9], elliptical trajectories of the system's motion are constructed based on a given central configuration, and the cases of systems of N = 4 and N = 5 particles with equal masses are also studied. The finiteness of the state of relative equilibrium of the system in the four-body problem was proven in [10]. In [11], a number of results were obtained for the more complex case of N = 5 particles in the system. The main mathematical aspects of this problem, are also considered in [12]. The known problem of rings (a small particle under the influence of a flat ring structure with equal masses and a central body) is considered in [13]. Systems of N particles with equal masses, which are located at the vertices of regular polygons, were considered in [14]. In [15], the problem of stability is studied for the symmetric case of a system of N particles on a plane. The existence of symmetrical structures in the form of central configurations of the system was proven in [16]. In [6], along with the spatial problem, the case is considered when particles of the system with the same masses are at the vertices of a regular *N*-gon in the problem on the plane and at the vertices of a regular polyhedron in the spatial problem, plus a particle with a small mass. Periodic solutions in the problem on the plane were also found in [17]. In [18], a problem on a plane was considered, when all the particles are on a circle and the center of mass of the system coincides with the center of the circle.

Along with analytical methods, various numerical methods are also used for this problem. The main difficulties encountered in modeling the motion of systems are associated with the large dimension of the problem as N increases and the non-regularity when particles approach each other. A parallel numerical method for this problem is presented in [19]. The work [20] examines the comparative effectiveness of various numerical methods for the problem of rings on a plane. Numerical iterative methods for the four-body problem on a plane were also proposed in [21]. [22] presents a number of basic algorithms for numerical simulation in the N-body problem, both for situations with possible collisions and for problems without particle collisions, and also gives an overview of the history of modeling of a systems in this problem.

Next, in Section 2 we consider the spatial problem of finding the minimum of the kinetic energy function of a system on its phase space under conditions on the size and angular momentum of a system. Based on the solution of this problem in Theorem 1, we prove further that the minimum possible kinetic energy of a system at the current value of its size can only be achieved on flat trajectories of a system. And under some natural additional conditions on the values of size and angular momentum of a system, this minimum can be achieved only on flat finite and periodic elliptical trajectories generated by some flat central configurations.

In Section 3 we consider a similar optimization dual problem of finding the minimum of the size of a system under conditions on its kinetic energy and angular momentum. Based on this optimization dual problem, a simpler solution is then obtained in Theorems 3 and 4, which is that the minimum possible size of a system at the current value of its kinetic energy can also be achieved only on flat trajectories. In this Section we also prove that under some additional conditions on the values of angular momentum and kinetic energy of a system, this minimum can also be achieved only on flat periodic elliptical trajectories generated by flat central configurations.

Section 4 considers the more complex spatial problem of finding local minima of the kinetic energy function of a system on its phase space at fixed values of the integrals of motion: angular momentum and total energy of a system. Based on the solution of this problem in Theorem 5, we prove further that under some natural additional conditions the local minima of the system kinetic and potential energy functions at fixed values of these integrals of motion can be achieved only on flat finite and periodic elliptical trajectories generated by some central configurations. And at the points of these local minima the minimum possible kinetic energy of the system at the current value of its size and the minimum possible size of the system at the current value of its kinetic energy are also achieved.

## 2. Minimum of the kinetic energy function of a system under conditions on its size and angular momentum

Consider a system of $N$ different material points (particles). Let $\boldsymbol{r} = (\boldsymbol{r}_1, \boldsymbol{r}_2, \ldots, \boldsymbol{r}_N) \in \Re$ be the complete set of radius-vectors of all particles of the system in three-dimensional space, $\boldsymbol{r}_i = (x_i, y_i, z_i)$ be the radius-vector of the particle with index $i = 1, \ldots, N$; where the coordinate space $\Re = \{\boldsymbol{r}: |\boldsymbol{r}_j - \boldsymbol{r}_i| > 0, \ i \neq j\} \subset E_{3N}$. (Here and below, indexes $i, j$ take the values $1, \ldots, N$; vectors are denoted in bold, $E_n$ is the $n$-dimensional Euclidean space.) We also denote the set of velocity vectors of various particles of the system as $\boldsymbol{v} = (\boldsymbol{v}_1, \boldsymbol{v}_2, \ldots, \boldsymbol{v}_N)$, where $\boldsymbol{v}_i = (\alpha_i, \beta_i, \gamma_i)$ is the velocity vector of the $i$-th particle; $\boldsymbol{v} \in \mho$, where $\mho$ is the set of possible values of $\boldsymbol{v} = (\boldsymbol{v}_1, \boldsymbol{v}_2, \ldots, \boldsymbol{v}_N)$, $\mho = E_{3N}$. The set (phase space) of all possible states $(\boldsymbol{r}, \boldsymbol{v})$ of a system of $N$ particles in this model has the form $\Omega = \Re \times \mho \subset E_{6N}$.

The trajectory of the system $[\boldsymbol{r}(t), \boldsymbol{v}(t)] \in \Omega$, $t \geq 0$ is determined by a system of differential equations with some initial conditions

$$m_i \dot{\boldsymbol{v}}_i = \boldsymbol{F}_i(\boldsymbol{r}), \quad i = 1, \ldots, N \tag{2.1}$$

$$\dot{\boldsymbol{r}}_i = \boldsymbol{v}_i, \quad i = 1, \ldots, N \tag{2.2}$$

Where the interaction force $\boldsymbol{F}_i(\boldsymbol{r})$ acting on the particle $i$ from all other particles of the system has the form

$$\boldsymbol{F}_i(\boldsymbol{r}) = \sum_{j \neq i} \boldsymbol{F}_{ij}(\boldsymbol{r}), \quad \boldsymbol{F}_{ij}(\boldsymbol{r}) = \frac{\gamma \, m_i \, m_j \, \boldsymbol{L}_{ij}(\boldsymbol{r})}{|\boldsymbol{L}_{ij}(\boldsymbol{r})|^3}, \tag{2.3}$$

where $\boldsymbol{F}_{ij}(\boldsymbol{r})$ is the interaction force acting on particle $i$ from particle j, $m_i > 0$ is the mass of particle $i = 1, \ldots, N$; $\gamma > 0$ is some constant, $\boldsymbol{L}_{ij}(\boldsymbol{r}) = \boldsymbol{r}_j - \boldsymbol{r}_i$, This model corresponds to the known $N$ – body problem, where formulas (2.3) correspond to Newton's law of gravity, when the force of attraction between two particles with masses $m_i, m_j$ is inversely proportional to the square of the distance between them.

A trajectory $[\boldsymbol{r}(t), \boldsymbol{v}(t)] \in \Omega$, $t \geq 0$ satisfying the system of differential equations (2.1), (2.2) under some initial conditions will be called stationary or finite (in terms of [23]), if there exist constants $0 < C_1 < C_2 < \infty$ such that $C_1 < |\boldsymbol{r}_j(t) - \boldsymbol{r}_i(t)| < C_2$ for each $i \neq j$, $t \geq 0$. And this trajectory will be called regular, if there exists a constant $0 < C < \infty$ such that $|r_j(t) - r_i(t)| > C$ for each $i \neq j, t \geq 0$.

Further we will assume that the origin of coordinates is at the center of mass of the system. As a characteristic of the size of the system in the state $(r, v) \in \Omega$ we take its "root-mean-square size" (further simply the size) $b = b(r) = \sqrt{\sum_i \beta_i |r_i|^2}$, where $\beta_i = m_i / \sum_j m_j$, $i = 1, \dots, N$. To simplify the formulas below, as a characteristic of the size of the system, we will also use the function $g = g(r) = \sum_i m_i |r_i|^2$, which is related to $b = b(r)$ by the relation $g = b^2 \sum_j m_j$. (Function $g(r)$ also has the meaning of the moment of inertia of the system in a problem on a plane.)

Let $L(r, v) = \sum_i m_i [r_i v_i]$ be the angular momentum of the system and $T(v) = \sum_i m_i |v_i|^2 / 2$ is the kinetic energy of the system in the state $(r, v) \in \Omega$. Consider the following problem for finding a minimum: It is required to find

$$min\ T(v),$$

where the minimum is taken over all states $(r, v) \in \Omega$ under conditions:

$$L(r, v) = L, \quad g(r) = g$$

Where $L = (L_x, L_y, L_z)$, $g > 0$ are some constants. Condition at the angular momentum of the system corresponds to three conditions on the components of this vector. We further assume without loss of generality that the vector $L$ is directed along the axis $z$, that is, $L = (0, 0, L)$ where $L = |L| > 0$. In a more detailed form in terms of variables $(x_i, y_i, z_i)$, $(\alpha_i, \beta_i, \gamma_i)$ this problem has the following form: it is required to find

$$min\ T(v) = (1/2)\ min \sum_i m_i (\alpha_i^2 + \beta_i^2 + \gamma_i^2) \tag{2.4}$$

where the minimum is taken over all states $(r, v) \in \Omega$ under the conditions

$$L_x(r, v) = \sum_i m_i (y_i \gamma_i - z_i \beta_i) = 0 \tag{2.5}$$

$$L_y(r, v) = \sum_i m_i (z_i \alpha_i - x_i \gamma_i) = 0 \tag{2.6}$$

$$L_z(r, v) = \sum_i m_i (x_i \beta_i - y_i \alpha_i) = L \tag{2.7}$$

$$g(r) = \sum_i m_i (x_i^2 + y_i^2 + z_i^2) = g \tag{2.8}$$

where the parameters of this problem $L > 0$, $g > 0$.

Next, in Theorems 1 and 2 we prove that the minimum possible kinetic energy of the system at the current value of its size (for a fixed value of the integral of motion: angular momentum $L$) can be achieved only on flat trajectories of the system. And under some additional conditions, these trajectories are flat finite and periodic (elliptic) trajectories generated by flat central configurations.

**Theorem 1.** *If at some point of the phase space $(r, v) \in \Omega$ the minimum in (2.4) is achieved under conditions (2.5) – (2.8), then at this point the equalities $z_i = 0$, $\gamma_i = 0$ are satisfied* for each $i = 1, \ldots, N$; that is, for all particles of the system, all vectors $\mathbf{r}_i = (x_i, y_i, z_i)$, $\mathbf{v}_i = (\alpha_i, \beta_i, \gamma_i)$ are in the same plane $z = 0$.

*Proof.* Let $B(L, g) \subset \Omega$ be a subset of the phase space over which the minimum in (2.4) is taken, that is, the set of all states $(\mathbf{r}, \mathbf{v}) \in \Omega$ for which conditions (2.5) – (2.8) are satisfied. And let $B^*(L, g) \subset \Omega^* = E_{6N}$ be the closure of this set (that is, it also includes points that satisfy these conditions, but at which the radius-vectors of different particles $\mathbf{r}_i, \mathbf{r}_j$ can coincide). Consider first the problem of finding the minimum (2.4) on the set $B^*(L, g)$. It is easy to show that this minimum is reached at some (perhaps not the only) point $(\mathbf{r}, \mathbf{v}) \in B^*(L, g)$. Indeed, let $B^*_C(L, g) \subset B^*(L, g)$ be a subset of all points that belong to $B^*(L, g)$ and such that $T(\mathbf{v}) \leq C$, where $C$ is any sufficiently large constant such that $C > \inf T(\mathbf{v})$ on the set $B^*(L, g)$. The set $B^*_C(L, g)$ is closed due to the continuity of the functions $L_x(\mathbf{r}, \mathbf{v}), L_y(\mathbf{r}, \mathbf{v}), L_z(\mathbf{r}, \mathbf{v}), g(\mathbf{r}), T(\mathbf{v})$ and is bounded, taking into account also the condition (2.8) at the function $g(\mathbf{r})$. Thus, the function $T(\mathbf{v})$ is continuous on a closed bounded set $B^*_C(L, g)$, whence it follows that the minimum in (2.4) is reached on the set $B^*_C(L, g)$. Consequently, this minimum is also reached on the set $B^*(L, g)$.

All functions in (2.4) – (2.8) have continuous partial derivatives at each point $(\mathbf{r}, \mathbf{v}) \in \Omega^* = E_{6N}$. The necessary conditions for the minimum in (2.4) on the set $B^*(L, g)$ at the point $(\mathbf{r}, \mathbf{v}) \in B^*(L, g)$ have the form:

$$\nabla T(\mathbf{v}) = a_1 \nabla L_x(\mathbf{r}, \mathbf{v}) + a_2 \nabla L_y(\mathbf{r}, \mathbf{v}) + a_3 \nabla L_z(\mathbf{r}, \mathbf{v}) - \lambda \nabla g(\mathbf{r}) \qquad (2.9)$$

Where $a_1, a_2, a_3, \lambda$ are some coefficients (Lagrange multipliers). Taking into account that $(\partial T / \partial \mathbf{r}_i) = \mathbf{0}, (\partial g / \partial \mathbf{v}_i) = \mathbf{0}, i = 1, \ldots, N$; equations (2.9) are further written as

$$\mathbf{0} = a_1 \frac{\partial L_x}{\partial \mathbf{r}_i} + a_2 \frac{\partial L_y}{\partial \mathbf{r}_i} + a_3 \frac{\partial L_z}{\partial \mathbf{r}_i} - \lambda \frac{\partial g}{\partial \mathbf{r}_i}, \quad i = 1, \ldots, N \qquad (2.10)$$

$$\frac{\partial T}{\partial \mathbf{v}_i} = a_1 \frac{\partial L_x}{\partial \mathbf{v}_i} + a_2 \frac{\partial L_y}{\partial \mathbf{v}_i} + a_3 \frac{\partial L_z}{\partial \mathbf{v}_i}, \quad i = 1, \ldots, N \qquad (2.11)$$

Equations (2.10) in variables $(x_i, y_i, z_i), (\alpha_i, \beta_i, \gamma_i)$ have the form:

$$2\lambda x_i = -a_2 \gamma_i + a_3 \beta_i, \quad i = 1, \ldots, N \qquad (2.12)$$

$$2\lambda y_i = a_1 \gamma_i - a_3 \alpha_i, \quad i = 1, \ldots, N \qquad (2.13)$$

$$2\lambda z_i = -a_1 \beta_i + a_2 \alpha_i, \quad i = 1, \ldots, N \qquad (2.14)$$

Equations (2.11) in variables $(x_i, y_i, z_i), (\alpha_i, \beta_i, \gamma_i)$, are written as

$$\alpha_i = a_2 z_i - a_3 y_i, \quad i = 1, \ldots, N \qquad (2.15)$$

$$\beta_i = -a_1 z_i + a_3 x_i, \quad i = 1, \ldots, N \qquad (2.16)$$

$$\gamma_i = a_1 y_i - a_2 x_i, \quad i = 1, \ldots, N \qquad (2.17)$$

Let's introduce the vector of coefficients $\mathbf{a} = (a_1, a_2, a_3)$. Then these equations can be written in a more compact vector form:

$$2\lambda \boldsymbol{r}_i = [\boldsymbol{v}_i \mathbf{a}], \quad \boldsymbol{v}_i = -[\boldsymbol{r}_i \mathbf{a}], \quad i = 1, \ldots, N \tag{2.18}$$

where $[\,\cdot\,]$ is the vector multiplication. Or in matrix form:

$$2\lambda \boldsymbol{r}_i = \|A\|\, \boldsymbol{v}_i, \quad \boldsymbol{v}_i = -\|A\|\, \boldsymbol{r}_i, \quad i = 1, \ldots, N \tag{2.19}$$

where the matrix $\|A\|$ has the form:

$$\|A\| = \begin{Vmatrix} 0 & a_3 & -a_2 \\ -a_3 & 0 & a_1 \\ a_2 & -a_1 & 0 \end{Vmatrix}$$

From equalities (2.18) it follows that at any minimum point $(\boldsymbol{r}, \boldsymbol{v}, a_1, a_2, a_3, \lambda)$, $(\boldsymbol{r}, \boldsymbol{v}) \in B^*(L, g)$ the vector $\mathbf{a} \neq 0$. Indeed, if $\mathbf{a} = 0$, then by virtue of the second equality in (2.18) at the minimum point the equalities $\boldsymbol{v}_i = 0$ are satisfied for each $i = 1, \ldots, N$; which contradicts the condition $L > 0$.

It also follows from (2.18) that at any minimum point the coefficient $\lambda \neq 0$. Indeed, let $\lambda = 0$. Then at the minimum point any non-zero vector $\boldsymbol{v}_i$ (at least one such vector exists due to condition $L > 0$) is collinear to the vector $\mathbf{a}$ due to the first equality in (2.18). But by virtue of the second equality in (2.18), this same vector is perpendicular to the vector $\mathbf{a}$, which means a contradiction. Thus, at any minimum point the coefficient $\lambda \neq 0$.

From (2.18) it also follows that at any minimum point $(\boldsymbol{r}, \boldsymbol{v}) \in B^*(L, g)$ for each particle of the system vectors $\boldsymbol{r}_i, \boldsymbol{v}_i$ are either simultaneously equal to zero or simultaneously not equal to zero. Note that if at the minimum point the state $(\boldsymbol{r}, \boldsymbol{v})$ of the system belongs to the phase space $\Omega$, that is $(\boldsymbol{r}, \boldsymbol{v}) \in B(L, g) \subset \Omega$, then equalities $\boldsymbol{r}_i = 0$, $\boldsymbol{v}_i = 0$ can be satisfied only for one (central) particle. From equations (2.18) it also follows that at any minimum point $(\boldsymbol{r}, \boldsymbol{v}) \in B^*(L, g)$ for each particle of the system with a non-zero radius vector $\boldsymbol{r}_i$, its velocity vector $\boldsymbol{v}_i$ is perpendicular to the radius vector.

Let's denote by $\mathbf{p_1} = (0, -a_3, a_2)$, $\mathbf{p_2} = (a_3, 0, -a_1)$, $\mathbf{p_3} = (-a_2, a_1, 0)$ the column vectors of the matrix $\|A\|$. Then, by virtue of (2.19), at any minimum point the following equalities hold:

$$2\lambda \boldsymbol{r}_i = \alpha_i \mathbf{p_1} + \beta_i \mathbf{p_2} + \gamma_i \mathbf{p_3}, \quad i = 1, \ldots, N \tag{2.20}$$

$$-\boldsymbol{v}_i = x_i \mathbf{p_1} + y_i \mathbf{p_2} + z_i \mathbf{p_3}, \quad i = 1, \ldots, N \tag{2.21}$$

For any coefficients $a_1, a_2, a_3$ the determinant of the matrix $\|A\|$ is equal to zero, i.e. this matrix is singular and, therefore, the vectors $\mathbf{p_1}, \mathbf{p_2}, \mathbf{p_3}$ are coplanar. Note that at the minimum point these vectors cannot be collinear. Indeed, then, by virtue of equalities (2.20), (2.21) the vectors $\boldsymbol{r}_i, \boldsymbol{v}_i$ for each particle are collinear (if these vectors are non-zero), which contradicts the condition $L > 0$. Thus, from (2.20), (2.21) it follows that at any minimum point $(\boldsymbol{r}, \boldsymbol{v}) \in B^*(L, g)$ all non-zero vectors $\boldsymbol{r}_i, \boldsymbol{v}_i$ are in the same plane spanned by coplanar vectors $\mathbf{p_1}, \mathbf{p_2}, \mathbf{p_3}$. And since, according to the conditions of the problem, the angular momentum vector of the system is directed along the axis $z$, this plane can only be the plane $z = 0$.

Thus, at any point of the minimum $(r, v) \in B^*(L, g)$ the following equalities hold: $z_i = 0$, $\gamma_i = 0$ for each $i = 1, \ldots, N$. Then the restrictions (2.5), (2.6) are satisfied automatically and become redundant, therefore $a_1 = a_2 = 0$. The necessary minimum conditions (2.12) – (2.17) for the vectors $r_i = (x_i, y_i)$, $v_i = (\alpha_i, \beta_i)$ on this plane are further written in the following form:

$$2\lambda x_i = a_3 \beta_i, \quad 2\lambda y_i = -a_3 \alpha_i, \quad i = 1, \ldots, N \tag{2.22}$$

$$\alpha_i = -a_3 y_i, \quad \beta_i = a_3 x_i, \quad i = 1, \ldots, N \tag{2.23}$$

Whence follows the equality $2\lambda = a_3^2$, after which the necessary minimum conditions take the form:

$$z_i = 0, \quad \gamma_i = 0, \quad i = 1, \ldots, N \tag{2.24}$$

$$\alpha_i = -a_3 y_i, \quad \beta_i = a_3 x_i, \quad i = 1, \ldots, N \tag{2.25}$$

Conditions (2.7), (2.8), taking into account these relations, are written as

$$\sum_i m_i (x_i \beta_i - y_i \alpha_i) = a_3 \sum_i m_i (x_i^2 + y_i^2) = L$$

$$\sum_i m_i (x_i \beta_i - y_i \alpha_i) = (1/a_3) \sum_i m_i (\alpha_i^2 + \beta_i^2) = L$$

$$\sum_i m_i (x_i^2 + y_i^2) = (1/a_3^2) \sum_i m_i (\alpha_i^2 + \beta_i^2) = g$$

from which follows the equality $a_3 = L/g$. From where we finally find that in problem (2.4) – (2.8) the necessary conditions for the minimum at point $(r, v) \in B^*(L, g)$ have the form:

$$z_i = 0, \quad \gamma_i = 0, \quad i = 1, \ldots, N \tag{2.26}$$

$$\alpha_i = -(L/g) y_i, \quad \beta_i = (L/g) x_i, \quad i = 1, \ldots, N \tag{2.27}$$

$$\sum_i m_i (x_i^2 + y_i^2) = g \tag{2.28}$$

Whence also it follows that the minimum of the function $T(v)$ in (2.4) on the set $B^*(L, g)$ is achieved at any point that satisfies the conditions (2.26) – (2.28), and this minimum is equal:

$$\min T(v) = L^2/(2g) \tag{2.29}$$

For some points $(r, v) \in B^*(L, g)$ that satisfy the necessary minimum conditions (2.26) – (2.28), the radius-vectors $r_i = (x_i, y_i)$ of different particles can coincide, which is impossible due to physical conditions in the phase space $\Omega$. Such points belong to the closure $B^*(L, g)$, but do not belong to the subset $B(L, g) \subset \Omega$, over which the minimum is taken in problem (2.4) – (2.8). But as can be seen from these relations, conditions (2.26) – (2.28) are also satisfied at some points $(r, v) \in B(L, g)$ (that is, at points with different vectors $r_i$). For this, taking into account (2.27), it is sufficient of the existence of non-coinciding vectors $r_i = (x_i, y_i)$ satisfying equality (2.28). Therefore, the minimum (2.29) is also achieved on the set $B(L, g) \subset \Omega$, from which, taking into account the necessary conditions for the minimum in (2.26), Theorem 1 follows. ∎

**Remark 2.1** As can be seen from the previous proof, at any point of the phase space $(r, v) \in B(L, g) \subset \Omega$ at which the minimum in the problem considered above is achieved, all particles of the system (except, may be, one central particle) have non-zero vectors $\boldsymbol{r}_i$, $\boldsymbol{v}_i$ and the velocity vector $\boldsymbol{v}_i$ of each particle is perpendicular to its radius vector $\boldsymbol{r}_i$.

**Remark 2.2** In the optimization problem considered above, generally speaking, the condition that the center of mass of the system is at the origin of coordinates is not used. However, Theorem 1 remains valid if we add one more constraint of the form $\sum_i m_i \boldsymbol{r}_i = 0$ in this problem, which takes into account this condition. Indeed, let

$$D = \{(r, v) \in \Omega: \sum_i m_i \boldsymbol{r}_i = 0\} \subset \Omega$$

be a subset on which this additional constraint holds. Then conditions (2.26) – (2.28) are obviously satisfied at some points from the subset $B(L, g) \cap D \subset B(L, g)$. For this, it is sufficient of the existence of points $(r, v) \in \Omega$ satisfying conditions (2.26), (2.27) and equalities $\sum_i m_i (x_i^2 + y_i^2) = g$, $\sum_j m_j x_j = 0$, $\sum_j m_j y_j = 0$. Whence it follows, taking into account the relation $B(L, g) \cap D \subset B(L, g)$ and equality (2.29), that in this optimization problem the minimum on the set $B(L, g) \cap D$ is also equal to the value (2.29) and this minimum is achieved only at points that satisfy the necessary conditions (2.26) – (2.28).

**Remark 2.3** Due to the necessary minimum conditions in (2.26) – (2.28), the trajectory of the system "starting" from any minimum point $(r, v) \in B(L, g) \subset \Omega$ is flat, but this trajectory, generally speaking, is not necessarily finite or regular depending on the initial point $(r, v)$.

In the following Theorem 2, we prove that the minimum possible value of the kinetic energy of the system at the current value of its size (for a fixed value of the integral of motion: angular momentum $L$) can only be achieved on flat trajectories of the system.

Let $H(L, g)$ be the solution of the optimization problem (2.4) – (2.8) considered above, where $L > 0$, $g > 0$ are the parameters of this problem, i.e.

$$H(L, g) = \min T(v),$$

where the minimum is calculated over all states $(r, v) \in B(L, g) \subset \Omega$. Then the function $H\{L, g[r(t)]\}$, $t \geq 0$ determines the lower bound for the value $T[v(t)]$ at the current value of the function $g[r(t)]$ on the regular trajectory $[r(t), v(t)] \in \Omega$ of the system for each $t \geq 0$. Where $L = |L| > 0$ is the value of angular momentum of the system on this trajectory, constant due to the known conservation laws ([1], [2], etc.). And taking into account the relation $b[r(t)] = \sqrt{g[r(t)]/\sum_j m_j}$, the function $H\{L, g[r(t)]\}$ also determines the lower bound for the kinetic energy $T[v(t)]$ of the system at the current value of its size $b[r(t)]$ for each moment of time $t \geq 0$.

**Theorem 2.** *If at some point of a regular trajectory $[r(t), v(t)] \in \Omega$, $t \geq 0$ the minimum possible kinetic energy $T[v(t)]$ of the system is achieved at the current value of its size $b = b[r(t)]$, then this trajectory can only be flat.*

*Proof.* From the definition of the function $H(L, g)$ it follows that on the trajectory $[\mathbf{r}(t), \mathbf{v}(t)] \in \Omega$, $t \geq 0$ the kinetic energy of the system $T[\mathbf{v}(t)]$ satisfies the inequality:

$$T[\mathbf{v}(t)] \geq H\{L, g[r(t)]\} \quad \text{for each} \quad t \geq 0$$

Where $g[\mathbf{r}(t)] = b^2 [\mathbf{r}(t)] \sum_j m_j$, $b[\mathbf{r}(t)]$ is the size of the system on the trajectory $[\mathbf{r}(t), \mathbf{v}(t)] \in \Omega$ at the moment of time $t \geq 0$, and the constant $L = |\mathbf{L}|$ is the value of the angular momentum of the system on this trajectory. But by the condition of the theorem on this trajectory the following equality holds:

$$T[\mathbf{v}(t)] = H\{L, g[r(t)]\} \quad \text{for some} \quad t \geq 0$$

Then the proof follows from the previous Theorem 1. ∎

**Additional conditions under which the minimum possible kinetic energy of the system is achieved on flat finite and periodic trajectories.** In accordance with Theorems 1 and 2, the minimum possible kinetic energy of the system at the current value of its size (for a fixed value of the integral of motion: angular momentum $L$) can only be achieved on flat trajectories of the system. Next, we will prove that under some additional conditions these trajectories are flat finite and periodic elliptical trajectories generated by flat central configurations. In addition, from the results of the Section 4 it will follow that at fixed values of the integrals of motion: angular momentum and total energy of a system the minimum possible kinetic energy of a system (in above sense) can be achieved only at the points of local minima of the functions $T(\mathbf{v})$, $f(\mathbf{r})$ of the system kinetic and potential energy.

Let $\mathbf{r} = (\mathbf{r}_1, \mathbf{r}_2, \ldots, \mathbf{r}_N) \in \Re_2$ be the complete set of radius vectors of all particles of the system on the plane, where $\mathbf{r}_i = (x_i, y_i)$ is the radius vector of the i-th particle, $\Re_2 = \{\mathbf{r} : |\mathbf{r}_j - \mathbf{r}_i| > 0, \ i \neq j\} \subset E_{2N}$ is the coordinate space of the system on the plane. And let $B_g = \{\mathbf{r} \in \Re_2 : g(\mathbf{r}) = g\} \subset \Re_2$ be a subset of all states $\mathbf{r} \in \Re_2$ of the system on this plane with a fixed value of the function $g(\mathbf{r}) = g$.

Let us denote by $\tilde{B}(L, g) \subset B(L, g) \subset \Omega$ the subset of points in the phase space that satisfy the necessary minimum conditions (2.26) – (2.28). In accordance with Theorem 1, the minimum of the function $T(\mathbf{v})$ on the set $B(L, g) \subset \Omega$ is achieved at each point $(\mathbf{r}^*, \mathbf{v}^*) \in \tilde{B}(L, g)$, where $\mathbf{r}^* \in B_g \subset \Re_2$. The trajectory of the system starting from point $(\mathbf{r}^*, \mathbf{v}^*) \in \tilde{B}(L, g) \subset \Omega$ is flat due to conditions (2.26) – (2.28), but this trajectory is not necessarily finite or regular depending on the initial point $\mathbf{r}^* \in B_g$. Let us denote by

$$f(\mathbf{r}) = (1/2) \sum_i \sum_{j \neq i} \frac{\gamma \, m_i \, m_j}{|\mathbf{r}_j - \mathbf{r}_i|} \tag{2.30}$$

the "force function", which coincides, up to a sign and an arbitrary constant, with the potential energy $U(\mathbf{r})$ of the system: $f(\mathbf{r}) = -U(\mathbf{r}) + C$ ([1], [2], etc.). Next, let us choose the initial point $\mathbf{r}^* \in B_g \subset \Re_2$ so that at this point the minimum of the potential energy function $f(\mathbf{r})$ of the system is achieved, that is, from the condition:

$$f(\mathbf{r}^*) = \min f(\mathbf{r}), \quad \mathbf{r} \in B_g \tag{2.31}$$

where the minimum is calculated over all states $r \in B_g$. Let's introduce also the function:

$$h(g) = \min f(r), \quad r \in B_g \tag{2.32}$$

where the minimum is calculated over all states $r \in B_g$. In accordance with [27], section 4 this function has the form:

$$h(g) = C(m)/\sqrt{g}, \quad g > 0 \tag{2.33}$$

where $C(m) > 0$ is a certain function of the masses of particles $m = (m_1, m_2, ..., m_N)$ of the system, for which the lower estimate holds ([27], section 5):

$$C(m) \geq \gamma \left( \sum_{i<j} m_i m_j \right)^{3/2} / \left( \sum_j m_j \right)^{1/2} \tag{2.34}$$

The minimum in (2.31), (2.32) is achieved at some (not necessarily unique) point $r^* = (r_1^*, \ldots, r_N^*) \in B_g \subset \Re_2$, at which the following necessary conditions for a minimum hold:

$$F_i(r^*) = -2\lambda m_i r_i^*, \quad i = 1, \ldots, N \tag{2.35}$$

where the coefficient (Lagrange multiplier) $\lambda > 0$ is related to the parameter $g$ by the relation: $g = [C(m)/2\lambda]^{2/3}$ ([27], sections 4 and 7). Next, we determine the velocity vectors $v^* = (v_1^*, \ldots, v_N^*)$ of the particles of the system at the initial point $(r^*, v^*)$ in accordance with (2.27) as follows: $v_i^* = (\alpha_i^*, \beta_i^*)$, where

$$\alpha_i^* = -(L/g) y_i^*, \quad \beta_i^* = (L/g) x_i^*, \quad i = 1, \ldots, N; \tag{2.36}$$

where $x_i^*, y_i^*$ are the coordinates of the radius vector $r_i^* = (x_i^*, y_i^*)$, $i = 1, \ldots, N$. Then, the point $(r^*, v^*)$ belongs to the subset $\tilde{B}(L, g) \subset B(L, g)$. That is, at this point the minimum of the function $T(v)$ on the set $B(L, g)$ is achieved. And by virtue of (2.35), (2.36), at this point the following equalities are satisfied:

$$F_i(r^*) = -2\lambda m_i r_i^*, \quad i = 1, \ldots, N \tag{2.37}$$

$$\alpha_i^* = -\omega_0 y_i^*, \quad \beta_i^* = \omega_0 x_i^*, \quad i = 1, \ldots, N \tag{2.38}$$

where $\omega_0 = L/g$, $\lambda = C(m)/(2g^{3/2})$. Relations (2.37), (2.38) define a flat trajectory of the type $\tilde{r}_\lambda(t, \omega_0)$ with parameters $\lambda, \omega_0$ (in the notation of [27]), generated by the flat central configuration $r^* \in B_g \subset \Re_2$ in (2.37). This trajectory is periodic and elliptic (in the sense that each particle of the system on this trajectory moves along an elliptic curve) with trajectory parameter $p = \omega_0^2/(2\lambda)$ and eccentricity $e = 1 - p$, if inequality $p < 2$ is satisfied ([27], section 8). That is, if the following inequality is satisfied for parameters $L, g$:

$$p = \omega_0^2/(2\lambda) = \frac{L^2}{C(m)\sqrt{g}} < 2 \tag{2.39}$$

which, taking into account relations (2.29), (2.33), is also equivalent to the following inequality:

$$T(v^*) < f(r^*), \qquad (2.40)$$

where $T(v^*)$, $f(r^*)$ are the values of the kinetic and potential energy of a system at the starting point $(r^*, v^*) \in \tilde{B}(L, g)$ of this trajectory. Note that the lower bound for the function $C(m)$ in (2.34) implies the following simpler sufficient condition for these inequalities in (2.39), (2.40) to hold:

$$L^2/\sqrt{g} < 2\gamma \left(\sum_{i<j} m_i m_j\right)^{3/2} / \left(\sum_j m_j\right)^{1/2} \qquad (2.41)$$

Thus, if inequalities (2.39), (2.40) or (2.41) are satisfied, then the minimum possible kinetic energy of the system at the current value of its size (for a fixed value of the integral of motion: angular momentum $L$) is achieved on a flat finite and periodic elliptic trajectory generated by the flat central configuration $r^* \in B_g$ (at which the minimum of the function $f(r)$ of the potential energy of the system on the set $B_g \subset \Re_2$ is achieved).

From (2.29), (2.33), (2.40) it also follows that on this trajectory the total energy of the system $E$ (constant due to the known conservation laws [1], [2], etc.) satisfies the following relations:

$$T(v^*) - f(r^*) = \frac{L^2}{2g} - \frac{C(m)}{\sqrt{g}} = E < 0 \qquad (2.42)$$

Next, from the results of the Section 4, taking into account the relations (2.39), (2.42), it will follow that at fixed values of the integrals of motion: angular momentum $L$ and total energy $E$ the minimum possible kinetic energy of a system at the current value of its size $b$ can be achieved only at the points of local minima of the functions $T(v)$, $f(r)$ of the system kinetic and potential energy (see further Remark 4.4).

## 3. Minimum of the system size under conditions on the kinetic energy and angular momentum of the system

Let's consider a problem that is dual ([24] – [26]) to the previous optimization problem (2.4) – (2.8) in Section 2: It is required to find

$$W(L, T) = \min g(r) = \min \sum_i m_i (x_i^2 + y_i^2 + z_i^2) \qquad (3.1)$$

where the minimum is calculated over all states $(r, v) \in \Omega$ under the conditions

$$L_x(r, v) = \sum_i m_i (y_i \gamma_i - z_i \beta_i) = 0 \qquad (3.2)$$

$$L_y(r, v) = \sum_i m_i (z_i \alpha_i - x_i \gamma_i) = 0 \qquad (3.3)$$

$$L_z(r, v) = \sum_i m_i (x_i \beta_i - y_i \alpha_i) = L \qquad (3.4)$$

$$T(v) = (1/2) \sum_i m_i (\alpha_i^2 + \beta_i^2 + \gamma_i^2) = T \qquad (3.5)$$

where the parameters $L = |L| > 0$, $T > 0$.

In the following Theorems 3 and 4 we prove that the minimum possible size of the system at the current value of its kinetic energy $T$ (for a fixed value of the integral of motion: the angular momentum $L$) can only be achieved on flat trajectories of the system. Note that the proof of these theorems is more concise compared to a similar result presented in Preprint [27], due to the use of the results of the previous section 2 for the optimization dual problem. In addition, in this section we also prove that under some additional conditions these trajectories are flat finite and periodic (elliptic) trajectories generated by flat central configurations.

**Theorem 3.** *If at some point of the phase space* $(r, v) \in \Omega$ *the minimum in (3.1) is achieved under conditions (3.2) – (3.5), then at this point the equalities* $z_i = 0$, $\gamma_i = 0$ *are satisfied for each* $i = 1, \ldots, N$; *that is, all vectors* $r_i = (x_i, y_i, z_i)$, $v_i = (\alpha_i, \beta_i, \gamma_i)$ *are in the same plane* $z = 0$.

*Proof* largely repeats the proof of the previous Theorem 1. The necessary minimum conditions in problem (3.1) – (3.5) can be written in the same form (2.9) – (2.11). After which, repeating relations (2.12) – (2.25), we find that the necessary conditions for the minimum in this problem have the following form:

$$z_i = 0, \quad \gamma_i = 0, \qquad i = 1, \ldots, N$$

$$\alpha_i = -a_3 y_i, \quad \beta_i = a_3 x_i, \quad i = 1, \ldots, N$$

Conditions (3.4), (3.5) are then written in the form of the following equalities:

$$\sum_i m_i (x_i \beta_i - y_i \alpha_i) = a_3 \sum_i m_i (x_i^2 + y_i^2) = L$$

$$\sum_i m_i (x_i \beta_i - y_i \alpha_i) = (1/a_3) \sum_i m_i (\alpha_i^2 + \beta_i^2) = L$$

$$\sum_i m_i (\alpha_i^2 + \beta_i^2) = a_3^2 \sum_i m_i (x_i^2 + y_i^2) = 2T$$

from which follows the equality $a_3 = (2T/L)$. From these relations we find that the complete set of necessary minimum conditions in problem (3.1) – (3.5) has the following form:

$$z_i = 0, \quad \gamma_i = 0, \qquad i = 1, \ldots, N \qquad (3.6)$$

$$\alpha_i = -(2T/L) y_i, \quad \beta_i = (2T/L) x_i, \quad i = 1, \ldots, N \qquad (3.7)$$

$$\sum_i m_i (x_i^2 + y_i^2) = L^2 / (2T) \qquad (3.8)$$

Whence also it follows that the minimum of the function $g(r)$ in this problem is equal to $L^2/(2T)$. After which the proof follows from the necessary conditions for the minimum in (3.6) – (3.8). ∎

**Theorem 4.** *If at some point of a regular trajectory* $[r(t), v(t)] \in \Omega$, $t \geq 0$ *the minimum possible size* $b = b[r(t)]$ *of the system is achieved at the current value of its kinetic energy* $T = T[v(t)]$, *then this trajectory can only be flat.*

*Proof* is similar to the previous Theorem 2. From the definition of the function $W(L, T)$ it follows that on the trajectory $[r(t), v(t)] \in \Omega$ the function $g[r(t)]$ satisfies the inequality:

$$g[r(t)] \geq W\{L, T[v(t)]\} \quad \text{for each} \quad t \geq 0$$

where $T[\boldsymbol{v}(t)]$ is the kinetic energy of the system at the moment of time $t \geq 0$ and constant $L = |\boldsymbol{L}|$ is the value of the angular momentum of the system on this trajectory. But by the condition of the theorem, taking into account also that $b\,[r(t)] = \sqrt{g[r(t)]/\sum_j m_j}$, on this trajectory the following equality holds:

$$g[\boldsymbol{r}(t)] = W\{L, T[\boldsymbol{v}(t)]\} \quad \text{for some} \quad t \geq 0$$

Then the proof follows from the previous Theorem 3. ∎

**Additional conditions under which the minimum possible size of the system is achieved on flat finite and periodic trajectories.** In accordance with Theorems 3 and 4, the minimum possible size $b$ of a system at the current value of its kinetic energy $T$ (for a fixed value of the integral of motion: the angular momentum $L$) can be achieved only on flat trajectories. Next, we will prove that under some additional conditions for the values of $L$, $T$ this minimum is achieved on flat finite and periodic elliptical trajectories generated by flat central configurations. In addition, from the results of the next Section 4 it will follow that at fixed values of the integrals of motion: angular momentum and total energy of a system the minimum possible size $b$ of a system (in above sense) can be achieved only at the points of local minima of the functions $T(\boldsymbol{v})$, $f(\boldsymbol{r})$ of the system kinetic and potential energy.

Let $B(L,T) \subset \Omega$ be a subset of the phase space on which the minimum in problem (3.1) – (3.5) is calculated, that is, the set of all points $(\boldsymbol{r}, \boldsymbol{v}) \in \Omega$ for which conditions (3.2) – (3.5) are satisfied. And let $\tilde{B}(L,T) \subset B(L,T)$ be the subset on which the necessary conditions (3.6) – (3.8) for the minimum in this problem are satisfied. In accordance with Theorem 3, the minimum of the function $g(\boldsymbol{r})$ on the set $B(L,T) \subset \Omega$ is equal to the value $g = L^2/(2T)$ and is achieved at each point $(\boldsymbol{r}^*, \boldsymbol{v}^*) \in \tilde{B}(L,T) \subset B(L,T)$. And by virtue of condition (3.7), at each minimum point $(\boldsymbol{r}^*, \boldsymbol{v}^*) \in \tilde{B}(L,T)$, the vector $\boldsymbol{v}^* = (\boldsymbol{v}_1^*, \ldots, \boldsymbol{v}_N^*)$ of particle velocities is uniquely determined by the vector of their coordinates $\boldsymbol{r}^* = (\boldsymbol{r}_1^*, \ldots, \boldsymbol{r}_N^*) \in B_g \subset \Re_2$ on the plane, where

$$B_g = \{\boldsymbol{r} \in \Re_2 : g(\boldsymbol{r}) = g\} \subset \Re_2, \quad g = L^2/(2T).$$

The trajectory of the system starting from the minimum point $(\boldsymbol{r}^*, \boldsymbol{v}^*) \in \tilde{B}(L,T) \subset \Omega$ is flat due to conditions (3.6) – (3.8), but is not necessarily finite or regular depending on the initial point $\boldsymbol{r}^* \in B_g \subset \Re_2$, where $g = L^2/(2T)$. Next, we define the initial point $\boldsymbol{r}^* \in B_g$, $g = L^2/(2T)$ similarly to the previous Section 2 so that at this point the minimum of the function $f(\boldsymbol{r})$ of the system potential energy is achieved on this subset, that is, from the condition:

$$h(g) = \min f(\boldsymbol{r}) = f(\boldsymbol{r}^*), \quad \boldsymbol{r} \in B_g \qquad (3.9)$$

where the function $f(\boldsymbol{r})$ is defined above in (2.30), and the minimum is calculated over all states $\boldsymbol{r} \in B_g \subset \Re_2$ with a fixed value of $g = L^2/(2T)$. In accordance with [27], section 4 the solution to this problem satisfies the above relation (2.33): $h(g) = C(\boldsymbol{m})/\sqrt{g}$ for each $g > 0$, where the function $C(\boldsymbol{m}) > 0$ satisfies inequality (2.34). And any point $\boldsymbol{r}^* \in B_g$ of the minimum in (3.9) satisfies the following equalities (necessary conditions for the minimum):

$$\boldsymbol{F}_i(\boldsymbol{r}^*) = -2\lambda m_i \boldsymbol{r}_i^*, \quad i = 1, \ldots, N \tag{3.10}$$

where $\lambda = C(\boldsymbol{m})/(2g^{3/2}) > 0$, $g = L^2/(2T)$. Let us determine the vector $\boldsymbol{v}^* = (\boldsymbol{v}_1^*, \ldots, \boldsymbol{v}_N^*)$ of particle velocities at the initial point $(\boldsymbol{r}^*, \boldsymbol{v}^*) \in \tilde{B}(L, T)$ of the trajectory, taking into account relations (3.7), as follows: $\boldsymbol{v}_i^* = (\alpha_i^*, \beta_i^*)$, where

$$\alpha_i^* = -(2T/L) y_i^*, \quad \beta_i^* = (2T/L) x_i^*, \quad i = 1, \ldots, N \tag{3.11}$$

where $x_i^*, y_i^*$ are the coordinates of the radius vector $\boldsymbol{r}_i^* = (x_i^*, y_i^*)$, $i = 1, \ldots, N$. Then point $(\boldsymbol{r}^*, \boldsymbol{v}^*)$ belongs to subset $\tilde{B}(L, T) \subset B(L, T)$. And by virtue of (3.10), (3.11), the following equalities are valid at this point:

$$\boldsymbol{F}_i(\boldsymbol{r}^*) = -2\lambda m_i \boldsymbol{r}_i^*, \quad i = 1, \ldots, N \tag{3.12}$$

$$\alpha_i^* = -\omega_0 y_i^*, \quad \beta_i^* = \omega_0 x_i^*, \quad i = 1, \ldots, N \tag{3.13}$$

where $\omega_0 = 2T/L$, $\lambda = C(\boldsymbol{m})/(2g^{3/2})$, $g = L^2/(2T)$. Relations (3.12), (3.13) define a flat trajectory of the form $\tilde{r}_\lambda(t, \omega_0)$ with parameters $\lambda, \omega_0$ (in the notation of [27]), generated by the flat central configuration $\boldsymbol{r}^* \in B_g$, $g = L^2/(2T)$ in (3.12). This trajectory is periodic and elliptic (in above sense) with eccentricity $e = 1 - p$, if the inequalities $0 < p < 2$ are satisfied, where the trajectory parameter $p = \omega_0^2/(2\lambda)$, ([27], section 8).

Thus, if the parameters $L, T$ satisfy inequality:

$$p = \omega_0^2/(2\lambda) = L\sqrt{2T}/C(\boldsymbol{m}) < 2 \tag{3.14}$$

then the minimum possible root-mean-square size $b$ of the system at the current value of its kinetic energy $T$ (at a fixed value of the integral of motion: angular momentum $L$) is achieved at the initial point $(\boldsymbol{r}^*, \boldsymbol{v}^*) \in \tilde{B}(L, T)$ of a flat finite and periodic elliptic trajectory generated by a flat central configuration $\boldsymbol{r}^* \in B_g \subset \Re_2$, $g = L^2/(2T)$ in (3.12). From which, in particular, it follows that if $L\sqrt{2T}/C(\boldsymbol{m}) \cong 1$, then this trajectory is approximately "circular" (i.e., each particle moves in a circle with the center at the origin) and the minimum possible size $b$ of the system is achieved or almost achieved in every point of this trajectory.

Note that at point $(\boldsymbol{r}^*, \boldsymbol{v}^*) \in \tilde{B}(L, T)$ the following equalities are satisfied:

$$T(\boldsymbol{v}^*) = T, \quad f(\boldsymbol{r}^*) = C(\boldsymbol{m})/\sqrt{g} = C(\boldsymbol{m})\sqrt{2T}/L \tag{3.15}$$

Whence it follows that inequality (3.14) is also equivalent to the following inequality:

$$T(\boldsymbol{v}^*) < f(\boldsymbol{r}^*), \tag{3.16}$$

where $T(\boldsymbol{v}^*)$, $f(\boldsymbol{r}^*)$ are the values of the functions of the kinetic and potential energy of the system at the initial point $(\boldsymbol{r}^*, \boldsymbol{v}^*)$ of this trajectory. Taking into account the lower bound for function $C(\boldsymbol{m})$ in (2.34), inequality (3.14) or (3.16) is valid if the following simpler sufficient condition is satisfied:

$$L\sqrt{T} < \sqrt{2}\gamma \left(\sum_{i<j} m_i m_j\right)^{3/2} / \left(\sum_j m_j\right)^{1/2} \tag{3.17}$$

Thus, if the parameters $L$, $T$ satisfy inequalities (3.14), (3.16) or the sufficient condition (3.17), then the minimum possible size $b$ of the system at the current value of its kinetic energy $T$ (at a fixed value of the integral of motion: angular momentum $L$) is achieved on a flat finite and periodic elliptic trajectory generated by a flat central configuration $r^* \in B_g$, on which the minimum of the function $f(r)$ of the system potential energy is achieved on the set $B_g \subset \Re_2$, where $g = L^2/(2T)$.

From the previous formulas (3.15), (3.16) it also follows that on this trajectory the total energy $E$ of the system (constant due to the known conservation laws [1], [2], etc.) satisfies the following relations:

$$T(v^*) - f(r^*) = T - \frac{C(m)\sqrt{2T}}{L} = E < 0, \tag{3.18}$$

Then from the results of the Section 4, taking into account the relations (3.14), (3.18), it will follow that at fixed values of the integrals of motion: angular momentum $L$ and total energy $E$ the minimum possible size $b$ of a system at the current value of its kinetic energy $T$ can be achieved only at the points of local minima of the functions $T(v), f(r)$ of the system kinetic and potential energy (see further Remark 4.4).

## 4. Local minima of the functions of the kinetic and potential energy of the system at fixed values of its angular momentum and total energy

Let $E(r, v) = T(v) - f(r)$ is the total energy of the system in state $(r, v) \in \Omega$, where $f(r)$ is the system potential energy function defined above in (2.30). Consider the spatial problem of finding local minima of the function $T(v)$ of the kinetic energy of a system on its phase space under the following conditions on the angular momentum $L(r, v)$ and total energy $E(r, v)$ of the system:

$$L(r, v) = L, \quad E(r, v) = E$$

where $L \neq 0$, $E < 0$ are some constants. Just as above, we will assume that the angular momentum vector $L$ of the system is directed along the axis $z$, that is, $L = (0, 0, L)$, where $L = |L| > 0$. In more detail, this problem has the following form: it is required to find

$$\text{local minima } T(v) \tag{4.1}$$

where the local minima are calculated over all states $(r, v) \in \Omega$ under the conditions:

$$L_x(r, v) = \sum_i m_i(y_i \gamma_i - z_i \beta_i) = 0 \tag{4.2}$$

$$L_y(r, v) = \sum_i m_i(z_i \alpha_i - x_i \gamma_i) = 0 \tag{4.3}$$

$$L_z(r, v) = \sum_i m_i(x_i \beta_i - y_i \alpha_i) = L \tag{4.4}$$

$$T(v) - f(r) = E \tag{4.5}$$

where constants $L > 0$, $E < 0$ are the parameters of this problem. If $N > 2$, then the global minimum $T(v)$ under the conditions (4.2) – (4.5) is not achieved (namely, $\inf T(v) = 0$). But in this problem, the downward convex function $T(v)$ can have quite a lot of conditional local minima, depending on the system parameters $N, m_1, m_2, \ldots, m_N$. Note also that if at some point $(r, v) \in \Omega$ of the phase space there is a local minimum of the kinetic energy function $T(v)$ of the system under conditions (4.2) – (4.5), then by virtue of condition (4.5) at this point there is also a local minimum of the potential energy function $f(r)$ of the system under the same conditions.

Next, we will prove that local minima of the functions $T = T(v)$, $f = f(r)$ at fixed values of the integrals of motion: angular momentum $L$ and total energy $E$ can be achieved only on flat trajectories of the system generated by flat central configurations. And under some additional conditions on the values of $L, E$ these trajectories are finite and periodic elliptic trajectories. In addition, taking into account the results of previous Sections 2 and 3, at the points of the local minima of the system characteristics $T, f$ the minimum possible kinetic energy of the system at the current value of its size and the minimum possible size of the system at the current value of its kinetic energy are also achieved.

**Theorem 5.** *If at some point $(r, v) \subset \Omega$ of the phase space a local minimum of the function $T(v)$ is achieved under conditions (4.2) – (4.5), then at this point the equalities $z_i = 0$, $\gamma_i = 0$ are satisfied for each $i = 1, \ldots, N$; that is, all vectors $r_i = (x_i, y_i, z_i)$, $v_i = (\alpha_i, \beta_i, \gamma_i)$, $i = 1, \ldots, N$ are in the same plane $z = 0$.*

*Proof.* Let $B(L, E) \subset \Omega$ be a subset of all states $(r, v) \in \Omega$ for which conditions (4.2) – (4.5) are satisfied. All functions $L_x(r,v)$, $L_y(r,v)$, $L_z(r,v)$, $T(v)$, $f(r)$, $E(r,v)$ have continuous partial derivatives at each point $(r, v) \in \Omega$. Necessary conditions for a local minimum of the function $T(v)$ under the conditions (4.2) – (4.5) at point $(r, v) \in B(L, E)$ have the form:

$$\nabla T(v) = a_1 \nabla L_x(r,v) + a_2 \nabla L_y(r,v) + a_3 \nabla L_z(r,v) + h \nabla E(r,v)$$

where $a_1, a_2, a_3, h$ are some coefficients. Or

$$(1 - h)\nabla T(v) = a_1 \nabla L_x(r,v) + a_2 \nabla L_y(r,v) + a_3 \nabla L_z(r,v) - h \nabla f(r)$$

Whence, taking into account that $\left(\partial f / \partial v_i\right) = 0$, $\left(\partial T / \partial r_i\right) = 0$, $i = 1, \ldots, N$; the equations follow:

$$h \frac{\partial f}{\partial r_i} = a_1 \frac{\partial L_x}{\partial r_i} + a_2 \frac{\partial L_y}{\partial r_i} + a_3 \frac{\partial L_z}{\partial r_i}, \quad i = 1, \ldots, N \tag{4.6}$$

$$(1 - h) \frac{\partial T}{\partial v_i} = a_1 \frac{\partial L_x}{\partial v_i} + a_2 \frac{\partial L_y}{\partial v_i} + a_3 \frac{\partial L_z}{\partial v_i}, \quad i = 1, \ldots, N \tag{4.7}$$

Equations (4.6) in the variables $(x_i, y_i, z_i)$, $(\alpha_i, \beta_i, \gamma_i)$, taking into account that $\left(\partial f / \partial r_i\right) = F_i(r)$, have the following form:

$$hF_{ix}(r) = m_i(-a_2\gamma_i + a_3\beta_i), \quad i = 1, \ldots, N \tag{4.8}$$

$$hF_{iy}(r) = m_i(a_1\gamma_i - a_3\alpha_i), \quad i = 1, \ldots, N \tag{4.9}$$

$$hF_{iz}(r) = m_i(-a_1\beta_i + a_2 \propto_i), \quad i = 1,\ldots,N \tag{4.10}$$

where $F_{ix}(r)$, $F_{iy}(r)$, $F_{iz}(r)$ are the coordinates of vector $F_i(r)$. Equations (4.7) in the variables $(x_i, y_i, z_i)$, $(\propto_i, \beta_i, \gamma_i)$ are written as

$$(1-h)\propto_i = a_2 z_i - a_3 y_i, \quad i = 1,\ldots,N \tag{4.11}$$

$$(1-h)\beta_i = -a_1 z_i + a_3 x_i, \quad i = 1,\ldots,N \tag{4.12}$$

$$(1-h)\gamma_i = a_1 y_i - a_2 x_i, \quad i = 1,\ldots,N \tag{4.13}$$

As above in the proof of Theorem 1, we introduce the vector of coefficients $\mathbf{a} = (a_1, a_2, a_3)$. Then these equations can be written in a more compact vector form:

$$hF_i(r) = m_i[v_i a], \quad (1-h)v_i = -[r_i a], \quad i = 1,\ldots,N \tag{4.14}$$

Or in matrix form:

$$hF_i(r) = m_i\|A\| v_i, \quad (1-h)v_i = -\|A\| r_i, \quad i = 1,\ldots,N \tag{4.15}$$

where matrix $\|A\|$ is given above in section 2.

Further, similarly to section 2, it follows from equations (4.14) that at the point of any local minimum $(r, v, a_1, a_2, a_3, h)$, $(r, v) \in B(L, E)$ the vector $\mathbf{a} = (a_1, a_2, a_3) \neq \mathbf{0}$. Indeed, let us prove this by contradiction: let $\mathbf{a} = \mathbf{0}$. Then, if $h \neq 1$, then by virtue of the second equation in (4.14) all vectors $v_i = \mathbf{0}$, which contradicts the condition $L > 0$. If $\mathbf{a} = \mathbf{0}$, $h = 1$, then by virtue of the first equation in (4.14) all vectors $F_i(r)$ are equal to zero, which is impossible at any point in the phase space at $N > 1$. Thus, at the point of any local minimum $\mathbf{a} \neq \mathbf{0}$.

Let us also prove that at the point of any local minimum, inequalities $h \neq 0$, $h \neq 1$ are satisfied. Indeed, if $h = 0$, $h \neq 1$, then by virtue of the first equation in (4.14) all velocity vectors $v_i$ are equal to zero or collinear to the vector $\mathbf{a}$. But then, by virtue of the second equation in (4.14), all non-zero vectors $v_i$ (at least one such vector can be found due to condition $L > 0$) are perpendicular to the vector $\mathbf{a}$, which means a contradiction.

If $h \neq 0$, $h = 1$, then by virtue of the second equation in (4.14) all non-zero radius-vectors $r_i$ are collinear to vector $\mathbf{a}$, which means that all vectors $F_i(r)$ are also collinear to vector $\mathbf{a}$. But then, by virtue of condition $L > 0$, there is at least one non-zero vector $v_k$, which is not collinear to vector $\mathbf{a}$, and by virtue of the first equation in (4.14), vector $F_k(r)$ is not equal to zero and is perpendicular to vector $\mathbf{a}$, which also means a contradiction.

If $h = 0$, $h = 1$, then by virtue of equations (4.14) at the local minimum point all non-zero vectors $r_i, v_i$ are collinear to vector $\mathbf{a}$, which also contradicts condition $L > 0$. Thus, at the point of any local minimum the inequalities $\mathbf{a} = (a_1, a_2, a_3) \neq \mathbf{0}$, $h \neq 0$, $h \neq 1$ are satisfied.

Let us further introduce, in the same way as above in section 2, the column vectors $\mathbf{p_1} = (0, -a_3, a_2)$, $\mathbf{p_2} = (a_3, 0, -a_1)$, $\mathbf{p_3} = (-a_2, a_1, 0)$ of the matrix $\|A\|$. Then, by virtue of (4.15), at the point $(r, v, a_1, a_2, a_3, h)$, $(r, v) \in B(L, E)$ of any local minimum the following equalities must hold:

$$hF_i(r) = m_i(\alpha_i \, \mathbf{p_1} + \beta_i \, \mathbf{p_2} + \gamma_i \mathbf{p_3}), \quad i = 1, \ldots, N \tag{4.16}$$

$$(h - 1)v_i = x_i \, \mathbf{p_1} + y_i \, \mathbf{p_2} + z_i \mathbf{p_3}, \quad i = 1, \ldots, N \tag{4.17}$$

For any coefficients $a_1, a_2, a_3$ the determinant of the matrix $\|A\|$ is equal to zero and this matrix is singular and, therefore, the vectors $\mathbf{p_1}, \mathbf{p_2}, \mathbf{p_3}$ are coplanar. And at the point of any local minimum these vectors cannot be collinear. Indeed, then, by virtue of equalities (4.16), (4.17), all vectors $F_i(r)$, $v_i$, $i = 1, \ldots, N$ are also collinear, which contradicts the first equation in (4.14), (taking into account also that at least one vector $v_i$ is non-zero due to condition $L > 0$). Thus, at the point of any local minimum, all non-zero vectors $F_i(r)$, $v_i$, $i = 1, \ldots, N$ are in the same plane spanned by coplanar vectors $\mathbf{p_1}, \mathbf{p_2}, \mathbf{p_3}$. It is not difficult to further show that if all vectors $F_i(r)$, $i = 1, \ldots, N$ are in the same plane, then all radius-vectors $r_i$, $i = 1, \ldots, N$ are also in this plane. Thus, by virtue of (4.16), (4.17), at the point of any local minimum all vectors $r_i$, $v_i$, $i = 1, \ldots, N$ are in the same plane spanned by coplanar vectors $\mathbf{p_1}, \mathbf{p_2}, \mathbf{p_3}$. And since, according to the conditions of the problem, the angular momentum vector of the system is directed along the axis z, this plane can only be the plane $z = 0$, which proves Theorem 5. ∎

**Remark 4.1** If at some point $(r, v) \in B(L, E) \subset \Omega$ there is a local minimum of the function $T(v)$ on the set $B(L, E) \subset \Omega$, then, by virtue of condition (4.5), at this point there is also a local minimum of the function $f(r)$ on this set.

Next, based on Theorem 5, we will prove that local minima of the functions $T = T(v)$, $f = f(r)$ of the system kinetic and potential energy (at fixed values of $L, E$) can only be achieved on flat trajectories of the system generated by certain flat central configurations. And under some additional conditions on the parameters $L, E$ these trajectories are flat and periodic elliptic trajectories. In addition, at the points of the local minima of the system characteristics $T, f$ the minimum possible kinetic energy of the system at the current value of its size and the minimum possible size of the system at the current value of its kinetic energy are also achieved (see further Remark 4.4).

Let at some point $(r, v) \in B(L, E)$ there is a local minimum of the function $T(v)$ on the set $B(L, E) \subset \Omega$. Then, in accordance with the proof of the previous Theorem 5, at the point $(r, v, a_1, a_2, a_3, h)$, $(r, v) \in B(L, E)$ the local minimum of the function $T(v)$ is achieved in problem (4.1) – (4.5), where $\mathbf{a} = (a_1, a_2, a_3) \neq \mathbf{0}$, $h \neq 0$, $h \neq 1$. And by virtue of Theorem 5, at this point all vectors $r_i$, $v_i$, $i = 1, \ldots, N$ are in the same plane z = 0. Then restrictions (4.2), (4.3) in this problem are satisfied automatically and become redundant, therefore, the coefficients $a_1 = a_2 = 0$. Necessary conditions (4.8) – (4.13) of the local minimum for the coordinates of vectors $r_i = (x_i, y_i)$, $v_i = (\alpha_i, \beta_i)$ on this plane take the following form:

$$hF_{ix}(r) = m_i a_3 \beta_i, \quad hF_{iy}(r) = -m_i a_3 \alpha_i, \quad i = 1, \ldots, N$$

$$(h - 1)\alpha_i = a_3 y_i, \quad (h - 1)\beta_i = -a_3 x_i, \quad i = 1, \ldots, N$$

where $r = (r_1, r_2, \ldots, r_N) \in \Re_2$. From where the equalities follow:

$$\boldsymbol{F}_i(\boldsymbol{r}) = -2\lambda m_i \boldsymbol{r}_i, \quad i = 1, \ldots, N \tag{4.18}$$

$$\alpha_i = \omega_0 y_i, \quad \beta_i = -\omega_0 x_i, \quad i = 1, \ldots, N \tag{4.19}$$

Where $\lambda = \omega_0^2(h-1)/(2h)$, $\omega_0 = a_3/(h-1) \neq 0$ and coefficients $\lambda, \omega_0$ must be determined depending on the parameters $L, E$ of the system. Note, that from these relations, in particular, it follows that at the point $(\boldsymbol{r}, \boldsymbol{v})$ of any local minimum of the functions $T(\boldsymbol{v}), f(\boldsymbol{r})$ on the set $B(L, E) \subset \Omega$, the equalities $\sum_i m_i \boldsymbol{r}_i = 0$, $\sum_i m_i \boldsymbol{v}_i = 0$ are satisfied. Relations (4.18), (4.19) determine a flat trajectory of the form $\tilde{r}_\lambda(t, \omega_0)$ with parameters $\lambda, \omega_0$ (in notation [27]), generated by the flat central configuration in (4.18).

Thus, from Theorem 5 it follows that the local minima of the functions $T(\boldsymbol{v}), f(\boldsymbol{r})$ of the kinetic and potential energy of a system (at fixed values of its angular momentum $L$ and total energy $E$) can be achieved only on flat trajectories generated by flat central configurations. Next, we will prove that under some additional conditions on the parameters $L, E$ these trajectories are the flat finite and periodic elliptical trajectories.

Let $\boldsymbol{r} = (\boldsymbol{r}_1, \boldsymbol{r}_2, \ldots, \boldsymbol{r}_N) \in \Re_2$ be the complete set of radius vectors of all particles of the system on the plane, where $\boldsymbol{r}_i = (x_i, y_i)$ is the radius vector of the particle $i$ and $\Re_2 = \{\boldsymbol{r} : |\boldsymbol{r}_j - \boldsymbol{r}_i| > 0, i \neq j \} \subset E_{2N}$ is the coordinate space of the system on the plane. Consider the following problem: it is required to find

$$\min [f(\boldsymbol{r}) + \lambda g(\boldsymbol{r})], \quad \boldsymbol{r} \in \Re_2 \tag{4.20}$$

where the minimum is calculated over all $\boldsymbol{r} \in \Re_2$. It is not difficult to show that for any $\lambda > 0$ the minimum in (4.20) is achieved at some (not necessarily unique) point $\boldsymbol{r}^* = (\boldsymbol{r}_1^*, \ldots, \boldsymbol{r}_N^*) \in \Re_2$. The necessary conditions for the minimum in (4.20) at this point have the following form:

$$\boldsymbol{F}_i(\boldsymbol{r}^*) = -2\lambda m_i \boldsymbol{r}_i^*, \quad i = 1, \ldots, N \tag{4.21}$$

At the minimum point $\boldsymbol{r}^* \in \Re_2$ the following relations are satisfied ([27], section 4):

$$g(\boldsymbol{r}^*) = (C/2\lambda)^{2/3}, \quad f(\boldsymbol{r}^*) = C^{2/3}(2\lambda)^{1/3} \tag{4.22}$$

where $C = C(\boldsymbol{m}) > 0$ is a function of $\boldsymbol{m} = (m_1, m_2, \ldots, m_N)$, for which the following lower estimate ([27], section 5) already used above, holds:

$$C(\boldsymbol{m}) \geq \gamma \left( \sum_{i<j} m_i m_j \right)^{3/2} / \left( \sum_j m_j \right)^{1/2} \tag{4.23}$$

(Note that this lower estimate gives the exact value of the function $C(\boldsymbol{m})$ for cases $N = 2$ and $N = 3$, [27], sections 5, 6). Next, we determine the vector $\boldsymbol{v}^* = (\boldsymbol{v}_1^*, \ldots, \boldsymbol{v}_N^*)$ of particle velocities at point $\boldsymbol{r}^* = (\boldsymbol{r}_1^*, \ldots, \boldsymbol{r}_N^*) \in \Re_2$, taking into account the relations (4.19), as follows: $\boldsymbol{v}_i^* = (\alpha_i^*, \beta_i^*)$, where

$$\alpha_i^* = \omega_0 y_i^*, \quad \beta_i^* = -\omega_0 x_i^*, \quad i = 1, \ldots, N \tag{4.24}$$

where $x_i^*, y_i^*$ are the coordinates of the radius vector $r_i^* = (x_i^*, y_i^*)$, $i = 1,\ldots,N$. Then, by virtue of (4.21), (4.24), point $(r^*, v^*) \in \Omega$ satisfies the necessary conditions (4.18), (4.19) for a local minimum of the function $T(v)$ on the set $B(L, E) \subset \Omega$, that is

$$F_i(r^*) = -2\lambda m_i r_i^*, \quad i = 1,\ldots,N \qquad (4.25)$$

$$\alpha_i^* = -\omega_0 y_i^*, \quad \beta_i^* = \omega_0 x_i^*, \quad i = 1,\ldots,N \qquad (4.26)$$

where the coefficients $\lambda > 0$, $\omega_0 \neq 0$ must be determined depending on the parameters $L, E$. Relations (4.25), (4.26) define a flat trajectory of the form $\tilde{r}_\lambda(t, \omega_0)$ with parameters $\lambda, \omega_0$, generated by the flat central configuration $r^* = (r_1^*, \ldots, r_N^*) \in \Re_2$ in (4.25), with the initial point $(r^*, v^*)$. This trajectory is periodic and elliptic with eccentricity $e = 1 - (\omega_0^2/2\lambda)$, if inequalities $0 < (\omega_0^2/2\lambda) < 2$ are satisfied ([27], section 8). Note that, due to relations (4.25), (4.26), at the initial point $(r^*, v^*) \in \Omega$ of this trajectory the following equalities are satisfied:

$$\sum_i m_i r_i^* = 0, \quad \sum_i m_i v_i^* = 0$$

and these equalities are preserved not only at the initial point, but also at all other points of this trajectory. From relations (4.22), (4.26) after simple transformations it follows that the characteristics of the system at the initial point $(r^*, v^*)$ of this trajectory are expressed through the parameter $p = \omega_0^2/(2\lambda)$ of this trajectory as follows:

$$f(r^*) = C^{2/3} \omega_0^{2/3}/p^{1/3} \qquad (4.27)$$

$$T(v^*) = \sum_i m_i (v_i^*)^2/2 = \omega_0^2 \, g(r^*)/2 = f(r^*)p/2 \qquad (4.28)$$

$$L(r^*, v^*) = \sum_i m_i |v_i^*||r_i^*| = \omega_0 \, g(r^*) = C^{2/3} p^{2/3}/\omega_0^{1/3} \qquad (4.29)$$

where $C = C(m)$. Next, we determine the coefficients $\lambda, \omega_0$ depending on the values of the system characteristics $L, E$ from condition $(r^*, v^*) \in B(L, E)$, i.e. from the following equalities:

$$L(r^*, v^*) = L, \quad T(v^*) - f(r^*) = E \qquad (4.30)$$

Whence, taking into account the previous formulas (4.27) – (4.29), we find that equations (4.30) have the form:

$$\frac{[C(m)]^{2/3} p^{2/3}}{\omega_0^{1/3}} = L, \quad \frac{[C(m)]^{2/3} \omega_0^{2/3}}{2p^{1/3}} (2 - p) = -E$$

From where we get the following equation for parameter $p = \omega_0^2/(2\lambda)$:

$$p(2 - p) = -2EL^2/[C(m)]^2 \qquad (4.31)$$

Let's introduce parameter $w = -2EL^2/[C(m)]^2$. If the system characteristics $L$, $E$ satisfy the inequalities:

$$E < 0, \quad 2|E|L^2 < [C(m)]^2, \qquad (4.32)$$

then equation (4.31) has two solutions:

$$p_1 = 1 - \sqrt{1-w}, \quad p_2 = 1 + \sqrt{1-w} \qquad (4.33)$$

where $0 < p_1 < p_2 < 2$. Thus, if the system characteristics $L, E$ are satisfied conditions (4.32), then local minima of the functions $T = T(v)$, $f = f(r)$ of the kinetic and potential energy of a system (at fixed values of angular momentum $L$ and total energy $E$) are achieved on two flat finite and periodic elliptic trajectories of the form $\tilde{r}_\lambda(t, \omega_0)$, generated by a flat central configuration in (4.25) with symmetrical eccentricity parameters:

$$e = 1 - p = \pm\sqrt{1-w} = \pm\sqrt{1 - \frac{2|E|L^2}{[C(m)]^2}} \qquad (4.34)$$

If conditions (4.32) for parameters $L, E$ are not satisfied, then equation (4.31) does not have a real solution, or local minima are achieved on trajectories on which the system is destroyed (particles of the system go to an infinite distance from each other). Note that the lower estimate for the function $C(m)$ on the right side in (4.23) allows us to find the following simpler sufficient condition for the fulfillment of these conditions for a system with particle masses $m = (m_1, m_2, \ldots, m_N)$:

$$0 < -2EL^2 < \gamma^2 \left(\sum_{i<j} m_i m_j\right)^3 / \sum_j m_j \qquad (4.35)$$

Next, from relations (4.29), (4.30) we find the parameters $\lambda, \omega_0$ of these two trajectories depending on the values of the system characteristics $L, E$:

$$\lambda_1 = p_1^3 [C(m)]^4 / (2L^6), \quad \omega_{01} = p_1^2 [C(m)]^2 / L^3$$

$$\lambda_2 = p_2^3 [C(m)]^4 / (2L^6), \quad \omega_{02} = p_2^2 [C(m)]^2 / L^3$$

Note that the initial parameters $a_3, h$ of the problem are related to the parameters $\omega_0, \lambda$ by the following relations: $a_3 = \omega_0/(p-1)$, $h = p/(p-1)$. Where, by virtue of (4.31), inequalities $p \neq 0$, $p \neq 1$ hold, if inequalities (4.32) are satisfied. Thus, this replacement of parameters is correct if conditions (4.32) are satisfied for the values of $L, E$.

From relations (4.27) – (4.30) we also find the values of local minima of the functions $T = T(v)$, $f = f(r)$ of the kinetic and potential energy of a system (at fixed values of its angular momentum $L$ and total energy $E$), which, in accordance with the previous relations, are achieved at the initial points of these two trajectories:

$$T_1 = p_1^2 [C(m)]^2 / (2L^2), \qquad f_1 = p_1 [C(m)]^2 / L^2 \qquad (4.36)$$

$$T_2 = p_2^2 [C(m)]^2 / (2L^2), \qquad f_2 = p_2 [C(m)]^2 / L^2 \qquad (4.37)$$

where the parameters $p_1, p_2$ are defined above in (4.33). Note that the values of the characteristic $g = g(r^*)$ of the system at these points, taking into account formulas (4.22), have the following form:

$$g_1 = p_1^{-2} L^4 / [C(m)]^2, \qquad g_2 = p_2^{-2} L^4 / [C(m)]^2 \qquad (4.38)$$

From (4.36) – (4.38) the equalities also follow:

$$g_1 = L^2/2T_1, \quad g_2 = L^2/2T_2 \tag{4.39}$$

which corresponds to the solution of problem (3.1) – (3.5) considered above in Section 3. That is, at the points of local minima of the system characteristics $T, f$ (at fixed $L, E$), the minimum possible root-mean-square size $b = \sqrt{g/\sum_j m_j}$ of the system is also achieved at the current value of its kinetic energy.

Thus, if conditions (4.32) or (4.35) are satisfied, then local minima of the functions $T = T(v)$, $f = f(r)$ of the kinetic and potential energy of a system (at fixed values of its angular momentum $L$ and total energy $E$) are achieved at the initial points of the two flat periodic elliptic trajectories considered above (with symmetrical parameters of the eccentricity: $e = \pm\sqrt{1-w}$) generated by flat central configurations in (4.25), (4.26).

**Remark 4.2** Due to symmetry relations (4.33), (4.34), these trajectories are actually two versions of the same periodic trajectory, which start from two different (opposite) points on this trajectory. Namely, the first trajectory starts from a point with angular coordinate $\varphi = 0$, and the second starts from a point of the same trajectory with angular coordinate $\varphi = \pi$. In other words, the local minima of the system characteristics $T, f$ (at fixed values of the integrals of motion $L, E$) considered above are achieved on the same (for example, the first) flat periodic trajectory of type $\tilde{r}_\lambda(t, \omega_0)$ with parameter $p_1 = \omega_{01}^2/(2\lambda_1) = 1 - \sqrt{1-w}$ and eccentricity:

$$e_1 = 1 - p_1 = \sqrt{1 - \frac{2|E|L^2}{[C(m)]^2}} \tag{4.40}$$

And local minima of the system characteristics $T, f$ (at fixed values of $L, E$) are achieved at points of this trajectory with angular coordinates $\varphi = 2l\pi$, $l = 0,1,2,\ldots$ with characteristics $T_1, f_1, g_1$ and at points with opposite angular coordinates $\varphi = (2l+1)\pi$, $l = 0,1,2,\ldots$ with characteristics $T_2, f_2, g_2$. At these same points, the minimum possible size $b$ of the system (in above sense) is also achieved (see further Remark 4.4).

From the previous relations it is also clear that if the parameter $w = -2EL^2/[C(m)]^2$ is close to 1, then the local minima of the system characteristics $T, f$ (at fixed values of $L, E$) and the minimum possible system size are achieved or almost achieved at all points of this approximately "circular" trajectory.

Note that relations of the form (4.21) are valid not only for the point $r^* \in \Re_2$ of the global minimum on the set $\Re_2 \subset E_{2N}$ in problem (4.20), but also for the point $r_k^* = (r_{k1}^*, \ldots, r_{kN}^*) \in \Re_2$ of any $k$th local minimum, that is

$$F_i(r_k^*) = -2\lambda m_i r_{ki}^*, \quad i = 1,\ldots,N; \quad k = 1,\ldots,n \tag{4.41}$$

Therefore, the above relations are also valid for the case of any $k$th local minimum in problem (4.20), if we replace the function $C(m)$ in these relations with a function $C_k(m)$ such that the following inequalities are satisfied ([27], section 7):

$$C_k(\boldsymbol{m}) \geq C(\boldsymbol{m}) \geq \gamma\left(\sum_{i<j} m_i m_j\right)^{3/2} / \left(\sum_j m_j\right)^{1/2}, \quad k = 1, \ldots, n \qquad (4.42)$$

where $n$ is the number of local minima on the set $\Re_2$ in problem (4.20). (The same is also true for the results of the previous Sections 2 and 3.)

Repeating the previous relations, we find that if the characteristics of the system $L, E$ satisfy the inequalities:

$$E < 0, \quad 2|E|L^2 < [C_k(\boldsymbol{m})]^2, \qquad (4.43)$$

then local minima of the system characteristics $T, f$ (at fixed values of the integrals of motion $L, E$) are achieved on a flat finite and periodic elliptic trajectory of type $\tilde{r}_\lambda(t, \omega_0)$, generated by the flat central configuration in (4.41), with trajectory parameter $p_k = 1 - \sqrt{1 - w_k}$, where $w_k = 2|E|L^2/C_k^2(\boldsymbol{m})$, and eccentricity:

$$e_k = 1 - p_k = \sqrt{1 - \frac{2|E|L^2}{[C_k(\boldsymbol{m})]^2}}, \quad k = 1, \ldots, n \qquad (4.44)$$

And these local minima of the system characteristics $T, f$ are achieved at points of this trajectory with angular coordinates $\varphi = 2l\pi, \ l = 0,1,2,\ldots$ and at points with angular coordinates $\varphi = (2l+1)\pi, \ l = 0,1,2,\ldots$ At these same points, the minimum possible size $b$ of the system at the current value of its kinetic energy $T$ is also achieved. If the parameter $w_k = -2EL^2/C_k^2(\boldsymbol{m})$ is close to 1, then the local minima of the characteristics $T, f$ (at fixed values of $L, E$) and the minimum possible system size $b$ (in above sense) are achieved or almost achieved at all points of this approximately circular trajectory.

**Remark 4.3** Since $C_k(\boldsymbol{m}) \geq C(\boldsymbol{m}), \ k = 1, \ldots, n$; then condition (4.35) is still a sufficient condition for the validity of the inequalities in (4.43) and formulas (4.44).

Thus, if the parameters $L, E$ satisfy conditions (4.43) or the simpler sufficient condition (4.35), then the local minima of the system characteristics $T, f$ at fixed values of the integrals of motion $L, E$ are achieved on $n$ flat periodic elliptical trajectories, generated by flat central configurations in (4.41), at points of these trajectories with angular coordinates $\varphi = 2l\pi$ and $\varphi = (2l+1)\pi, \ l = 0,1,2,\ldots$ . At these same points of these trajectories, the minimum possible size b of the system (in above sense) is also achieved. In the general case, the number of such points of local minima of the characteristics $T, f$ (at fixed values of $L, E$) is obviously equal to $2q$, where $q \leq n$ is the number of indices k for which the main conditions (4.43) for the parameters $L, E$ are satisfied.

**Remark 4.4** Equations (2.42) and (3.18) obtained above in Sections 2 and 3, after simple transformations are reduced to equation (4.31). Whence it follows that the solution of the problems (2.4) – (2.8) and (3.1) – (3.5) of the previous Sections, can only be achieved at the points of local minima of the functions $T = T(\boldsymbol{v}), \ f = f(\boldsymbol{r})$ of the system kinetic and potential energy (at fixed values of the integrals of motion $L, E$) considered above. Thus, at the points of these local minima of the system characteristics $T, f$ the minimum possible kinetic energy of the system at the current value of its size and the minimum possible size $b$ of the system at the current value of its kinetic energy are also achieved.

The previous relations depend on the functions $C(m), C_k(m)$ which in the general case are quite complex. However, the lower bound for these functions in (4.23), (4.42) allows us to obtain a simpler sufficient condition (4.35) under which local minima of the system characteristics $T, f$ (at fixed values of $L, E$) are achieved on flat finite and periodic elliptical trajectories, as well as estimates of the system characteristics at the points of these local minima.

**Remark 4.5** In the case of $N = 2$ and $N = 3$, the lower bound for the function $C(m)$ on the right side of (4.23) gives the exact value of this function ([27], sections 5 and 6). If $N = 2$, then from the formula on the right side of (4.23) the equality follows: $C(m) = \gamma m_1 m_2 \sqrt{m}$, where $m = m_1 m_2/(m_1 + m_2)$. In this case, the flat periodic elliptic trajectory with eccentricity (4.40) found above as a solution of the problem (4.1) – (4.5) (see also Remark 4.2), coincides with a well-known solution of the Kepler's two-body problem (see, for example, [23], par.15, etc.). Condition (4.32) for the existence of local minima of the system characteristics $T, f$ (at fixed values of the integrals of motion $L, E$) in this case, as can be seen from (4.31), (4.33), is equivalent to the condition: $0 < p < 2$ of the finiteness of this trajectory (in other words, the condition of the existence of the system).

## 5. Conclusion

Thus, in accordance with Theorems 1 and 2, the minimum possible kinetic energy of the system at the current value of its size can be achieved only on flat trajectories of the system. And under some additional conditions for the size and angular momentum of the system these trajectories are flat finite and periodic elliptical trajectories generated by some central configurations. In accordance with Theorems 3 and 4, a similar result hold, namely, the minimum possible size of the system at the current value of its kinetic energy can also be achieved only on flat trajectories of the system. And under additional conditions for the values of angular momentum and kinetic energy of the system, the minimum possible size of the system (in above sense) can be achieved only on flat periodic elliptical trajectories generated by flat central configurations.

In section 4 above it is considered also the spatial problem of finding local minima of the function of the kinetic energy of a system on its phase space at fixed values of the integrals of motion: angular momentum and total energy of a system. In accordance with Theorem 5 under some natural additional conditions on the values of these characteristics, the local minima of the system kinetic and potential energy functions (at fixed values of the integrals of motion) can be achieved only on flat finite and periodic elliptical trajectories generated by some flat central configurations. And at the points of these local minima the minimum possible kinetic energy of the system at the current value of its size and the minimum possible size of the system at the current value of its kinetic energy are also achieved.

Considering the physical meaning of the system kinetic and potential energy functions, it is natural to further assume that systems in a stable state move in space along the flat finite periodic trajectories considered above, on which local minima of these characteristics (at fixed integrals of motion) are achieved and the structure of the system is preserved under the conditions obtained above. This hypothesis provides a partial explanation at a qualitative level for some properties of planetary systems and galaxies, which still remain largely unclear. Namely, the existence of such huge systems as a single whole in space (which in itself is surprising), the conditions of existence (finiteness) and the flat structure of the system, etc.

In addition, numerical modeling using gradient descent methods for a number of system models with different parameters $N, m_1, m_2, \ldots, m_N$ shows the spiral structure of the system on the flat trajectories considered above (in this case, the distance of the system particles from the origin or center of mass, the greater, the smaller the particle mass). Considering the results presented above, such a hypothesis is rather natural, but nevertheless needs further additional arguments and research.

Igor Pavlov, Department of Mathematics, Bauman Moscow State Technical University, ul. Baumanskaya 2-ya, 5, Moscow, 105005, Russian Federation.

E-mail address: ipavlov@bmstu.ru , igorpavlov5@mail.ru , igvalpavlov@gmail.com